\documentstyle{article}

\begin{document}



\def\um{{1 \over 2} } 
\def\lh{\mbox{\boldmath $ \ell $}}
\def\lv{\vec \ell }
\def\xiv{\vec \xi }
\def\lie{{\pounds}_{\xiv}\, } 
\def\lg{\lie g(x) } 
\def\lm{\lie \eta } 
\def\ll{\lie \lh} 
\def\fd{\dot F } 
\def\fdd{\ddot F } 
\def\cd{{\dot C}_1 } 
\def\cdd{{\ddot C}_1 } 
\def\dd{{\dot D}_2 } 
\def\ddd{{\ddot D}_2 } 
\def\bd{{\dot B}_3 } 
\def\th{\mbox{\boldmath $ \Theta $}} 
\def\pfi{\partial_{\varphi} } 
\def\ks{{\em KS \/}} 
\def\gms{{\em GMS \/}} 
\def\gks{{\em GKS \/}} 
\def\ab{\alpha \beta} 

\newtheorem{defi}{Definition}
\newtheorem{theo}{Theorem}
\newtheorem{coro}{Corollary}
\newtheorem{prop}{Proposition}
\newtheorem{lem}{Lemma}[section]

\def\P{{\it Proof:} \hspace{3mm}}
\def\R{{\rm I\!R}}
\def\N{\hfill \rule{2.5mm}{2.5mm}}
\def\K{{\cal K}}
\def\Kl{{\cal K}_{\ell}}
\def\Cl{{\cal C}_{\ell}}


\title{Kerr{-}Schild Symmetries}
\author{Bartolom\'e Coll\footnotemark[5], Sergi R.
Hildebrandt\footnotemark[3], and Jos\'e M. M. Senovilla\thanks{Also at
Laboratori de F\'{\i}sica Matem\`atica,
Societat Catalana de F\'{\i}sica, IEC, Barcelona, Spain.} \footnotemark[4]\\ \\
\footnotemark[5] Syst\`emes de R\'ef\'erence Spatio-temporels,
Observatoire de Paris-CNRS UMR 8630,\\
61, avenue de l'Observatoire, Paris F-75014 France \\
\footnotemark[3] Institut d'Estudis Espacials de Catalunya, IEEC/CSIC,
Edifici Nexus 201,\\
Gran Capit\`a 2-4, 08034 Barcelona, Spain \\
\footnotemark[4] Departamento de F\'{\i}sica Te\'orica, Facultad de Ciencias,\\
Universidad del Pa\'{\i}s Vasco, Apartado 644, 48080, Bilbao, Spain.
}
\date{}
\maketitle

\begin{abstract}
We study continuous groups of generalized Kerr-Schild transformations 
and the vector fields that generate them in any $n$-dimensional manifold with a 
Lorentzian metric. We prove that all these vector 
fields can be intrinsically characterized and that 
they constitute a Lie algebra {\it if} the null deformation direction is 
fixed. The properties of these Lie algebras are briefly analyzed and 
we show that they are generically finite-dimensional but that they may 
have infinite dimension in some relevant situations. The most general 
vector fields of the above type are explicitly constructed for the 
following cases: any two-dimensional metric, the general spherically 
symmetric metric and deformation direction, and the flat metric with 
parallel or cylindrical deformation directions. 
\end{abstract}
\section{Introduction}
The classical Kerr-Schild Ansatz \cite{ks}, in which one considers 
metrics of the form $\tilde g = \eta + 2 H \lh \otimes \lh$, where 
$\eta$ is the Minkowski metric and $\lh$ is a null 1-form, was very
succesful in finding exact solutions of the vacuum Einstein field 
equations, and Kerr-Schild type of metrics had been studied 
before with other aims \cite{Tra}. As is well known, the celebrated 
Kerr metric was in fact originally presented in its Kerr-Schild form 
\cite{Kerr}, and the general Kerr-Schild vacuum solution was
explicitly found \cite{ks,ks2}. The Ansatz was also succesfully 
applied to the Einstein-Maxwell equations \cite{ks2,E-M} and to the case of null 
radiation \cite{null}. The Kerr-Schild metrics were also analyzed on 
theoretical grounds, see for instance \cite{varios}, and a review 
with the main results can be found in \cite{kramer}.

The Kerr-Schild Ansatz was soon generalized to the case in which the 
base metric is not {\it flat} \cite{Tho,taub,BG,NA}. Thus, two metrics
${\tilde g}$ and $g$ are linked by a {\em generalized Kerr-Schild 
relation} if there exist a function $H$ and a null 1-form $\lh$ such that
$$ {\tilde g} = g + 2 H \lh \otimes \lh . $$
A possible physical interpretation of this relation has been recently 
put forward in \cite{nota2'}.
Again, many exact solutions to Einstein's field equations have been 
found by using the generalized Kerr-Schild Ansatz. Several 
examples are given in \cite{varios2} for vacuum and Einstein-Maxwell 
and in \cite{ms} for perfect fluids. The general vacuum to vacuum 
generalized Kerr-Schild metric was also solved in \cite{gp}.

In this paper, we take a point of view which seems to have not been 
adopted hitherto, namely, that the above formula
is the deformation that a {\it transformation} of the spacetime 
produces on the metric, and we will simply use the term {\it Kerr-Schild 
transformation}. In this sense, Kerr-Schild transformations are on the 
same footing as isometries (which leave the metric invariant, 
$\tilde{g}=g$), or conformal transformations ($\tilde{g}=\Psi g$). As 
in the latter cases, in many situations the
interesting point is not the existence of a discrete transformation,
but the existence of a continuous group of such transformations
admitted by the given metric. This is our aim, so that we shall consider
continuous groups of Kerr-Schild transformations \cite{nota3},
or {\em Kerr-Schild groups}.
As in the case of Killing or conformal vector fields which generate
the afore-mentioned classical transformations, we shall show that such
groups are generated by what we call {\em Kerr-Schild vector fields}
$\xiv$, solutions to the equations
$$
\lie g = 2 h \lh \otimes \lh, \hspace{1cm}
\lie \lh = m \lh 
$$
and that they form a Lie algebra. However, the Kerr-Schild groups are 
associated to the metric structure of the spacetime $g$
{\em as well as} to a field of null directions $ \lh$. Among other 
implications of this fact, we shall prove that the Lie algebra 
of Kerr-Schild vector fields can be of infinite dimension \cite{nota4}.



The paper is organized as follows. In Section 2 Kerr-Schild
vector fields are introduced and the basic definitions are given. Section 3
is devoted to the general properties of such fields. We 
provide the general equations that they must satisfy {\it 
independently} of the null direction $\lh$, and make some 
considerations about the structure of the set of all such vector 
fields in a given spacetime. In particular, we also show that they form
a Lie algebra {\it for each} direction $\lh$, and that these Lie 
algebras can be infinite dimensional. In Section 4 we present 
several explicit examples: a) we find the general solution for the case 
of an arbitrary 2-dimensional metric, which depends on 4 arbitrary 
functions, two for each possible null direction; b) the solution for
the case of a parallel null direction in flat $n$-dimensional spacetime
is given, and shown to depend on $n$ arbitrary functions of one 
variable and on $(n-2)(n-3)/2$ arbitrary constants; c) the general 
solution for the case in which the metric as well as the deformation 
direction are spherically symmetric is explicitly found. Several cases 
appear and some well-known metrics arise naturally; and d) the case of 
a cylindrical deformation direction in flat spacetime is analyzed, 
with some surprising results about the local character of the solutions.
Finally, the last Section contains some conclusions and the possible 
lines for additional work.

\section{Kerr-Schild vector fields}
Let $(V_n,g)$ be an $n$-dimensional manifold with a metric
$g$ of Lorentzian signature (--,+,\dots,+). Indices in $V_n$ run from 0 to
$n-1$ and are denoted by Greek small letters. The tensor and exterior
products are denoted by $\otimes$ and $\wedge$, respectively, boldface letters
are used for 1-forms and arrowed symbols for vectors,
and the exterior differential is denoted by $d$. The pullback of any
application $\phi$ is $ \phi^{*}$ and the Lie derivative
with respect to the vector field $\xiv$ is written as $\lie$.
Equalities by definition are denoted by
$\equiv$, and the end of a proof is signalled by \rule{2.5mm}{2.5mm}.

\begin{defi}[Kerr{-}Schild group] A
one-parameter group of transformations $ \{ \phi_s \} $ of $ V_n $,
$ s\in \R $, is called a {\em Kerr{-}Schild group} \ if
the transformed metric is of the form
$$ \phi^{*}_s (g) = g + 2 H_s \lh \otimes \lh,$$
where $ \lh $ is a null 1-form field and $ H_s $ are
functions over $V_n$.
\end{defi}
It is easily seen that the group structure ($ \phi_{s} \phi_{r} = \phi_{s+r} $)
requires a transformation law of the form
$$     \phi^{*}_s (\lh) = M_s\, \lh $$
where $M_s$ is a function over $V_n$ for each $s$.

Let us denote by $\xiv $ the infinitesimal generator
of such a group. By writing $ h \equiv dH_s/ds |_{s=0} $,
$ m \equiv dM_s/ds |_{s=0}$, a standard calculation leads to
\begin{prop}[Kerr-Schild equations] The generator \ $\xiv $\ of a
Kerr-Schild group satisfies the equations
\begin{eqnarray}
\lie g = 2 h \, \lh \otimes \lh, \label{eq:KS1}\\
\lie \lh = m \, \lh \ \label{eq:KS2}
\end{eqnarray}
where $h$ and $m$ are two functions over $V_n$.
\N
\end{prop}

Note that the set of equations (\ref{eq:KS2}) is nothing
but the guarantee that the form of (\ref{eq:KS1}) is stable
under Lie derivatives of arbitrary order $p$, that is,
$(\lie )^{(p)} g = 2h^{(p)}\lh \otimes \lh \ $, where $h^{(p)} \equiv
\lie h^{(p-1)} +2 m h^{(p-1)}$. The usual results on differential
equations ensure that, conversely, any $\xiv$ satisfying
(\ref{eq:KS1}-\ref{eq:KS2}) generates a Kerr-Schild group which is
generically local, so that it will define a
{\em local  Kerr{-}Schild group of local transformations.}

It is convenient to know the contravariant
version of equations (\ref{eq:KS1}-\ref{eq:KS2}),
\begin{eqnarray*}
\lie g^{-1} = - 2 h \,  \lv \otimes \lv, \hspace{1cm}
 \lie \lv  \equiv [\xiv,\lv\,] = m \,  \lv ,
\end{eqnarray*}
and also their expressions with index notation
\begin{eqnarray*}
\nabla_{\alpha} \xi_{\beta} + \nabla_{\beta} \xi_{\alpha}= 2 h \,
\ell_{\alpha} \ell_{\beta},\hspace{15mm}
 \xi ^{\rho}\nabla_{\rho} \ell_{\alpha} + \ell _{\rho}\nabla_{\alpha}
\xi^{\rho} = m \, \ell_\alpha.
\end{eqnarray*}

As $\lh$\ is null, the function $h$ is not an
invariant of the tensor $2h\, \lh \otimes \lh,$ which can be equally
characterized by any other pair $h'= A^{-2}h ,\ \lh ' =
A\, \lh$, where $A$ is a non-vanishing $C^{\infty}$
function, so that equations (\ref{eq:KS2}) become
$\lie \lh ' = m'\, \lh '$, with \ $ m' = m + \lie \log |A| $.

It is worth to remark that, in contrast with the classical isometries
or conformal transformations, the Kerr-Schild groups take into account
the metric deformation $\lie g$ {\it with regard} to a given null
direction $\lh$. In this sense, and in order to be precise,
we give the following
\begin{defi}[Kerr-Schild vector fields] Any solution $\xiv$ of the
Kerr-Schild equations (\ref{eq:KS1}-\ref{eq:KS2}) will be called a
{\em Kerr-Schild vector field (KSVF)} with respect to the {\em deformation
direction} $\lh$. The functions $h$ and $m$ are the {\em gauges} of the
metric $g$ and of the deformation $\lh$, respectively.
\end{defi}
Obviously, any Killing vector field which leaves invariant the
deformation direction $\lh$ is also a Kerr-Schild vector field with
$h=0$. Thus, as is usual in similar contexts, we define
\begin{defi}[Proper Kerr-Schild vector fields]
A non-zero Kerr-Schild vector field $\xiv$ will be called {\em proper} if
its metric deformation $\lie g$ is non-vanishing.
\end{defi}
In other words, $\xiv \neq \vec{0}$ is a proper KSVF if the corresponding
metric gauge is non-zero, $h\neq 0$. The zero vector $\xiv =\vec{0}$ 
is also considered to be a proper KSVF for any deformation direction.

\section{General properties of Kerr-Schild vector fields}
A first, straightforward, property of Kerr-Schild vector fields is
\begin{prop} Two metrics related by a Kerr-Schild transformation,
$ {\tilde g} = g + 2 H \lh \otimes \lh$,
admit the same KSVFs with respect to $\lh$.
\label{same}
\N
\end{prop}
\begin{coro} Every KSVF $\xiv$ of a metric $g$ is a Killing
vector field of a Kerr-Schild transformed metric ${\tilde g}$ of $g$.
\end{coro}
\P For any related Kerr-Schild metric ${\tilde g}$, one has
$\lie {\tilde g}= 2 {\tilde h}\lh \otimes \lh$ with
${\tilde h}\equiv h + \lie H + 2mH$, and thus the equation
${\tilde h}=0$ admits local solutions in the unknown $H$.\N

Notice that this does not mean in general that the set of KSVFs is the
isometry algebra of some Kerr-Schild related metric, because the
solutions $H$ are in general different for each $\xiv$. The
conditions under which there exists a common solution $H$ for all
$\xiv$, that is to say, a new Kerr-Schild metric for which {\em
all} KSVFs are Killing fields, will be given elsewhere.

The natural question arises whether or not the set of {\it all} KSVFs
for a metric, regardless of their deformation directions, can be
characterized in some sense. This is answered in the following
\begin{theo}
A vector field $\xiv$ in $(V_n,g)$ is a KSVF for some deformation
direction if and only if
\begin{eqnarray}
\lie g \times \lie g =0, \label{uni1} \\
(\lie \lie g) \wedge \lie g = 0 .\label{uni2}
\end{eqnarray}
\label{uni}
\end{theo}
\P We are using the notation $(t\times T)_{\mu\nu}\equiv
t_{\mu\rho}T_{\nu}{}^{\rho}$ for the inner product of any two
rank-2 tensors $t$ and $T$. On the other hand, the second
equation (\ref{uni2}) simply means that there exists some function
$\Psi$ such that
\begin{equation}
\lie \lie g = \Psi \lie g \, .\label{uni2'}
\end{equation}
Now, if the Kerr-Schild equations (\ref{eq:KS1}-\ref{eq:KS2}) hold,
then it is very simple to check (\ref{uni1}-\ref{uni2}). Conversely,
assume that (\ref{uni1}-\ref{uni2}) are satisfied.
As is known (see e.g. \cite{PR}), any 2-index symmetric tensor $t$ with the
property $t\times t =0$ must have the form $t=2 h \lh \otimes \lh$ for some
null 1-form $\lh$ and some function $h$, so that from (\ref{uni1}) it
follows (\ref{eq:KS1}) at once. Using this, the equation (\ref{uni2'})
readily gives
$$
h\left(\lie \lh \otimes \lh +\lh \otimes \lie \lh \right)=
\left(h\Psi - \lie h\right)\lh \otimes \lh
$$
which leads to (\ref{eq:KS2}) if $h\neq 0$ or is empty if $h=0$.\N
\begin{coro}
If $\xiv$ is a proper KSVF of the metric $g$, then its deformation
direction can be explicitly constructed as
\begin{equation}
\lh ' \equiv \left|(\lie g)(\vec{u},\vec{u})\right|^{-1/2} i(\vec u)\lie g
\label{ele}
\end{equation}
where $\vec u$ is an {\em arbitrary} timelike vector.
\end{coro}
It is worthwhile to note that, despite what it might seem, expression
(\ref{ele}) for $\lh '$ is {\em independent} of $\vec{u}$. 
\P By $i(\vec u)T$ we mean the usual contraction
$(i(\vec u)T)_{\mu}=u^{\rho}T_{\rho\mu}$.  To
prove (\ref{ele}), assume that $\xiv$ is a proper KSVF, so that equations
(\ref{uni1}-\ref{uni2}) hold and we can set $\lie g =2 \epsilon \lh '
\otimes \lh '$ with $\epsilon =\pm 1$. Then, by contracting with
$\vec u$ once and twice, the expression (\ref{ele}) follows.\N

Theorem \ref{uni} allows to define a well-posed initial-value problem
for KSVFs by considering equations (\ref{uni2}) (or equivalently
(\ref{uni2'})) as the evolution system and the remaining set
(\ref{uni1}) as the constraint equations for the initial data
set. In fact, we have
\begin{coro}
The system of equations for the general KSVFs of $g$ is
involutive: if the equations (\ref{uni2'}) are satisfied in an open set
$\Omega \subseteq V_{n}$ and the equations (\ref{uni1}) hold on a
hypersurface $\Sigma \subset \Omega$ non-tangent to $\xiv$, then
(\ref{uni1}) are satisfied all over $\Omega$.
\end{coro}
\P Let us see the evolution of the constraint equations (\ref{uni1})
under the action of $\lie$. Note that
$$
\lie (\lie g \times \lie g)= (\lie \lie g) \times \lie g +
\lie g \times (\lie \lie g)- \lie g \times \lie g \times \lie g
$$
so that using (\ref{uni2'}) it follows
$$
\lie (\lie g \times \lie g)=2\Psi (\lie g \times \lie g)-
\lie g \times (\lie g \times \lie g)
$$
which proves the assertion, because this is a first order ODE for the 
constraint and its unique solution with zero initial condition 
vanishes.\N

It is interesting to remark that many other well-known sets of equations
are also constraints for (\ref{uni2'}), such as the cases
of Killing or conformal vector fields. In this sense,
the KSVFs satisfy a set of evolution equations which is common to
Killing or conformal vectors, and they differ from each other
in the constraints for the initial data. Furthermore, systems of the type
$$
(\lie)^{(p)} g =\Psi (\lie)^{(p-1)} g
$$
were considered some years ago by Papadopoulos \cite{Pap}, so that one
could say that the proper KSVFs of a metric are of Papadopoulos type with
$p=2$ constrained to satisfy the conditions (\ref{uni1}).

As is well-known, the Killing or the conformal vector fields provide 
constants of motion along geodesic curves (null geodesics for the conformal 
fields). As we are going to prove now, this also holds in an 
appropriate sense for the KSVFs. Let us remind first that a 
differentiable curve $\gamma$ is called a {\it subgeodesic with 
respect to the vector field $\vec{p}$} (see e.g. \cite{Scho})
if its tangent vector $\vec{v}$ satisfies 
$\nabla_{\vec{v}}\vec{v}=a\vec{v} + \lambda \vec{p}$ for some $a$ 
and $\lambda$. As with the case of geodesics curves, one can always 
choose an affine parametrization along $\gamma$ such that one can set 
$a=0$. For our purposes, only a subset of the subgeodesics are of interest.
\begin{defi}[$\ell$-parametrized subgeodesics]
Any subgeodesic with
$\lambda =(\ell_{\mu}v^{\mu})^2$
and an affine parametrization
will be called an affinely {\em $\ell$-parametrized subgeodesic}.
\end{defi}
Let us remark that this definition is given for arbitrary $\vec{p}$, 
and only the scalar $\lambda$ is restricted.
The affinely $\ell$-parametrized subgeodesics 
with tangent vector $\vec{v}$ are the solutions to the ordinary
differential equations
$$
\frac{dv^{\mu}}{d\tau}+\left(\Gamma^{\mu}_{\nu\rho}+
p^{\mu}\ell_{\nu}\ell_{\rho}\right)v^{\nu}v^{\rho}=0
$$
which only need as initial conditions the value of $\vec{v}$ at any 
given point. Then, a typical calculation of
$\nabla_{\vec{v}}(\xiv\cdot \vec{v})$ leads to
\begin{prop}
Let $\vec{v}$ be the tangent vector of an affinely $\ell$-parametrized 
subgeodesic $\gamma$ and $\xiv$ a KSVF with respect
to $\lh$ and metric gauge $h$. Then, $\xiv\cdot \vec{v}$ is constant 
along $\gamma$ whenever $\xiv\cdot \vec{p} + h=0$.\N
\end{prop}
Let us remark that the condition $\xiv\cdot \vec{p} + h=0$ is very 
weak in the sense that it is not very restrictive. For instance, for 
any proper KSVF and any field of directions $\vec{P}$ non-orthogonal to $\xiv$ 
the above condition simply fixes the appropriate factor which must 
multiply $\vec{P}$ to define the subgeodesics with respect to that 
direction. In other words, by simply choosing
$\vec{p}=-h \, \vec{P}/(\xiv\cdot \vec{P})$ the condition holds. 

Nevertheless, there are important differences between the classical
Killing or conformal vector fields and the KSVFs. To start with, the
set $\K$ of all KSVFs for a given metric does not have the structure of a
vector space, as is obvious from the non-linear character of the
relation (\ref{uni2'}) or directly from the Kerr-Schild equations if
several deformations directions are taken into account. However, one
can define $\Kl$ as the set of all KSVFs with regard to $\lh$.
Obviously, $\Kl = {\cal K}_{\ell '}$ for any other $\lh '=A\lh$, so
that the sensible thing to do is to consider $\Kl$ only for the direction
defined by $\lh$ or $\lh '$. To that end, let us denote by $\Cl$ the
null congruence of integral curves of $\lv$, (so that $\Cl \equiv
{\cal C}_{\ell '}$.) Then, the set $\K$ can be written as the union
$$
\K = \bigcup_{\Cl} \Kl .
$$
The interesting point here is that each of the $\Kl$ is a vector
space and, in fact, one has the following result.
\begin{prop} The set $\Kl$ of solutions $\xiv$ to the
equations (\ref{eq:KS1}-\ref{eq:KS2}) form a Lie algebra, hereafter
called the {\em Kerr-Schild algebra with respect to $\Cl$}.
\end{prop}
\P By construction, the set $\Kl$ has an evident {\em vector space}
structure because if $\xiv$ and $\vec \zeta $ are any two
KSVFs with regard to the same $\lh$, then any linear combination with
constant coefficients $c_{1}\xiv +c_{2} \vec\zeta$ is also a KSVF with
regard to the same deformation direction. Let us denote by
$h_{\vec \xi}$, $m_{\vec \xi}$ and $h_{\vec \zeta}$, $m_{\vec
\zeta}$ the gauge functions associated to $\xiv$ and
$\vec \zeta $, respectively. The identity
$$ {\pounds}_{[\xiv , \vec \zeta ]} =
\lie {\pounds}_{\vec \zeta } -{\pounds}_{\vec \zeta }\lie ,$$
applicable to any tensor field, immediately leads to
\begin{eqnarray*}
{\pounds}_{[\xiv , \vec \zeta ]}\, g =
2 h_{[\xiv , \vec \zeta ]}\, \lh \otimes \lh , \hspace{15mm}
{\pounds}_{[\xiv , \vec \zeta ]}\lh = m_{[\xiv , \vec \zeta ]}\, \lh ,
\end{eqnarray*}
with
\begin{eqnarray*}
h_{[\xiv , \vec \zeta ]}= \lie h_{\vec \zeta }-
{\pounds}_{\vec \zeta} h_{\xiv} +
2(h_{\vec \zeta }m_{\xiv} - h_{\xiv}m_{\vec \zeta }), \hspace{1cm}
m_{[\xiv , \vec \zeta ]} = \lie
m_{\vec \zeta }- {\pounds}_{\vec \zeta } m_{\xiv}
\end{eqnarray*}
where the Kerr-Schild equations (\ref{eq:KS1}-\ref{eq:KS2}) for
$\xiv$ and $\vec \zeta $ have been used.\N

It is interesting to observe that the equations
(\ref{eq:KS2}), which were {\em necessary} to ensure the
local group property in the {\em one-parameter case}, are  also {\em
sufficient} to produce the Lie algebra structure in the {\em multidimensional}
case.

Despite the above, the set $\K$ is {\it not} the direct sum of the $\Kl$'s for
all $\Cl$ because one has $\Kl \cap {\cal K}_{k}\neq \{\vec 0\}$ in
general. Still, $\K$ can be expressed as a simple direct sum sometimes
as follows from the following results.
\begin{lem}
If a KSVF belongs to two different Kerr-Schild algebras, then it is a
Killing vector field.
\end{lem}
\P If $\xiv \in \Kl \cap {\cal K}_{k}$ with $\Cl \neq {\cal C}_{k}$,
then $\lie g=2h \lh \otimes \lh =
2 f \mbox{\boldmath $ k $} \otimes \mbox{\boldmath $k$}$ with
$\mbox{\boldmath $k$} \wedge \lh \neq 0$, from where $h=f=0$.\N
\begin{prop}
The set $\K$ is the following disjoint union
$$
\K = \left[\bigsqcup_{\Cl} \hat{\Kl} \right]\sqcup Kil
$$
where $Kil$ is the Lie algebra of Killing vector fields in $(V_{n},g)$
and each $\hat{\Kl}$ is the subset of $\Kl$ formed by the proper
KSVFs with regard to $\Cl$. Thus, if there are no Killing vectors in 
the spacetime, the set $\K$ is the direct sum of Lie algebras
$$
Kil = \{\vec 0\} \hspace{1cm} \Longrightarrow \hspace{1cm} 
\K = \bigoplus_{\Cl} \hat{\Kl} = \bigoplus_{\Cl} \Kl .
$$
\end{prop}
\P The proof is immediate from the above Lemma, because $Kil\cap
\hat{\Kl}=\{\vec 0\}$ for all $\Cl$.\N

The above result does not say that we have a direct sum of {\it
finite-dimensional} Lie algebras. Of course, we know that $Kil$ is
always of finite dimension, but we do not know yet about the other
algebras $\Kl$. As we are going to prove now, they are {\it
generically} finite-dimensional, but there are some special
degenerate cases in which some of them are infinite-dimensional. First,
we need a Lemma identifying the cases when a KSVF leaves invariant
{\em every} single integral curve of $\Cl$.
\begin{lem} There are proper KSVFs $\xiv$ tangent to its deformation direction
$\lh$, $\mbox{\boldmath $\xi$} \wedge \lh = 0$, if and only if $\lv$ is
geodesic,  shear-free, expansion-free, and the 1-form $\mbox{\boldmath $a$}$
appearing in
\begin{equation}\label{eq:Ll}
 \pounds_{\lv}\, g =
\mbox{\boldmath $a$} \otimes \lh + \lh \otimes \mbox{\boldmath $a$}
\end{equation}
has the form
\begin{equation}\label{eq:da}
\mbox{\boldmath $a$} = -d \log |\mu | + \frac{h}{\mu} \lh
\end{equation}
for some functions $\mu,h$.
\label{lema}
\end{lem}
\P If $\xiv = \mu \lv$, the invariance of $\Cl$ is trivial, so that to prove
the Lemma only equations (\ref{eq:KS1}) must be checked. They readily
lead to (\ref{eq:Ll}), with $\mbox{\boldmath $a$}$ given
in (\ref{eq:da}). Equations (\ref{eq:Ll}) can be rewritten as
$$
\nabla_{\nu} \ell_{\mu}+\nabla_{\mu} \ell_{\nu}=
\ell_{\mu}a_{\nu}+\ell_{\nu}a_{\mu}
$$
which characterize the geodesic, shear-free and expansion-free null 1-forms
$\lh $, see e.g. \cite{kramer}.\N

From the well-known Goldberg-Sachs theorem and its generalizations
(see e.g. \cite{kramer}), the existence of geodesic and shear-free
null congruences is severely restricted in arbitrary spacetimes, so
that the possibility above is rather exceptional. Nevertheless, these
exceptions are of great interest, as they include many of the simpler
and/or physically relevant spacetimes, see the next section.


Now we can prove the
infinite-dimensional character of some of the Lie algebras $\Kl$.
\begin{theo} Two vector fields $\xiv $ and $\rho  \xiv $ with $d\rho \ne 0$
are KSVFs with respect to the same deformation direction $\lh$ if
and only if $\mbox{\boldmath $\xi$} \wedge \lh = 0$ and $\lh$ is
integrable $\lh \wedge d\lh = 0$ satisfying (\ref{eq:Ll}-\ref{eq:da}).
Then the functions $\rho$ are those of
the ring generating $\lh$, that is to say, such that
$ \lh \wedge d\rho = 0$.
\end{theo}
\P The Kerr-Schild equations (\ref{eq:KS1}) for both $\xiv$ and $\rho  \xiv $
imply that $\xiv = \mu\lv$ and $d\rho = \sigma \lh \neq 0$. The second of these
conditions implies that the null $\lh$ is irrotational (and therefore geodesic)
with $ \lh \wedge d\rho = 0$, while the first one says that we are in the
situation of Lemma \ref{lema}, so that (\ref{eq:Ll}-\ref{eq:da}) hold.
Conversely, if $\lh$ satisifies (\ref{eq:Ll}-\ref{eq:da}) then the vector field
$\xiv = \mu\lv$ is a KSVF with regard to $\lh$ and metric gauge $h$. 
As $\lh$ is also hypersurface-orthogonal we have $du\wedge \lh =0$ for
some non-vanishing funtion $u$. Then, for {\it any} function $\rho (u)$ we have
$$
\pounds_{(\rho\xiv)}\, g =\pounds_{(\rho\mu\lv)}\, g \propto \lh \otimes \lh
$$
which proves the result.\N

Notice that this result means that {\it all} the vector fields $\rho\mu\lv$
are KSVFs for {\it arbitrary} $\rho$, as long as $\rho$ is in the ring
generating $\lh$. That is to say, these KSVFs depend on an arbitrary function
$\rho (u)$, with $\lh \propto du$, and thus the corresponding Lie algebra
$\Kl$ has infinite dimension.
Other infinite-dimensional algebras associated to a metric
structure exist such as, for example, curvature collineations \cite{kld},
but they are directly related to the partly antisymmetric Riemann tensor
in degenerate cases, and not to the regular symmetric metric $g$, as in the
present case.
Explicit examples of infinite-dimensional Kerr-Schild algebras will
be presented in the next section.

\section{Explicit examples of Kerr-Schild vector fields}

In this Section, explicit expressions for the KSVFs in several situations
of relevance and interest are given. Some implications on
the corresponding (generalized) Kerr-Schild related metrics are
derived and briefly commented.

\subsection{General two-dimensional spacetime $(V_2,g)$}
\label{subsec:2dim}
The most general line-element for $n=2$ can be locally written as
\begin{equation}
ds^2=2 e^f du dv\, , \hspace{1cm} f=f(u,v) \label{2g}
\end{equation}
and there are only two inequivalent null directions given by
$du$ and $dv$. From a theoretical point of view,
it is enough to find the KSVFs associated
with the deformation direction $\lh = du$ (say), and then the solutions for 
the deformation direction $dv$ are analogous interchanging $u$ with $v$. 
We have
\begin{prop}
For any 2-dimensional metric, the most general KSVF with respect to $\lh = du$
is given by
\begin{equation}
\xiv = a(u)\frac{\partial}{\partial u} +
B\left(u,v;f;a(u),b(u)\right) \frac{\partial}{\partial v}
\label{2dim}
\end{equation}
where $a(u),b(u)$ are two arbitrary functions and $B$ is the general 
solution of
$$
\frac{\partial}{\partial v} \left(e^fB\right)=-\frac{\partial}{\partial u} 
\left(e^fa\right).
$$
The deformation and metric gauges are then given by
\begin{equation}
m=\dot{a}, \hspace{1cm} h=e^f\frac{\partial B}{\partial u}   
\label{2dimg}
\end{equation}
where a dot means derivative with respect to the argument.
\end{prop}
\P From (\ref{eq:KS2}) one easily gets
$$
\lie du = d(\lie u) = m du
$$
which fixes the component of $\xiv$ along $\partial /\partial u$ as 
an arbitrary function $a(u)$ and gives the first equation of (\ref{2dimg}). 
Now, the remaining Kerr-Schild equations (\ref{eq:KS1}) are equivalent 
to
$$
\lie \left(e^f dv\right) + \dot{a} e^f dv = h du
$$
which can be rewritten as
$$
dB=e^{-f}\left(h\, du -(\dot{a} + \lie f) dv\right).
$$
This is the desired result. Let us note that one only has to solve 
the part of the above equation giving the derivative
$\partial B/\partial v$, which depends on the arbitrary integrating 
function $b(u)$, and then $h$ is simply isolated as written in 
(\ref{2dimg}).\N

Thus, the solution in this case depends on {\it two} arbitrary 
functions of one variable $u$. This is an explicit example in which 
the Kerr-Schild algebra has infinite dimension. 

Similarly, one can derive the general 
solution for the other posible deformation direction $dv$, getting
\begin{equation}
\xiv = A\left(u,v;f;c(v),d(v)\right)\frac{\partial}{\partial u} +
d(v) \frac{\partial}{\partial v}
\label{2dim'}   
\end{equation}
where now $c$ and $d$ are arbitrary functions of $v$, and the 
corresponding metric gauges are $m=\dot d$ and
$h=e^f\partial A/\partial v$. These KSVFs are proper if and only 
if $\partial A/\partial v \neq 0$, and analogously for (\ref{2dim}),
so that we have also obtained the following result.
\begin{coro}
The set $\K$ of all KSVFs of any two-dimensional spacetime 
$(V_{2},g)$ can be written as the disjoint union
$$
\K = \hat{\K}_{du}\sqcup \hat{\K}_{dv} \sqcup Kil
$$
where $\hat{\K}_{du}$ is the set of all vector fields 
of the form (\ref{2dim}) with $\partial B/\partial u \neq 0$, 
$\hat{\K}_{dv}$ is the set of all vector fields of the form (\ref{2dim'})
with $\partial A/\partial v \neq 0$, and $Kil$ is the Lie algebra of 
Killing vector fields.\N
\end{coro}

Therefore, the set $\K$ can be completely and explicitly constructed 
for $n=2$ in general, and it depends on four arbitrary functions. This 
is the maximum freedom one can attain in two dimensions, so that the 
above Corollary suggests the validity of the following theorem, which
can be certainly proven.
\begin{theo}
Any two-dimensional Lorentzian $g$ is a Kerr-Schild 
transformed metric of the flat two-dimensional Minkowski metric.
\end{theo}
\P Starting with the general metric $g$ given by (\ref{2g}), one can 
set $d\tilde v\equiv -H du + e^f dv$ for some $H$ as long as the 
integrability conditions
$\partial H/\partial v =-e^f \partial f/\partial u$ hold. This has 
always solution for $H$, so that we have
$$
ds^2 = 2dud\tilde v + 2H du^2
$$
which is the desired result as $2dud\tilde v$ is obviously flat.\N

Obviously, the combination of this Theorem with Prop.\ref{same} allows 
to obtain the expressions (\ref{2dim},\ref{2dim'}) in a simple way.

\begin{coro}
Any pair of 2-dimensional Lorentzian metrics are related by a 
Kerr-Schild transformation with respect to any of the two possible 
null deformation directions.\N
\end{coro}
These nice simple results are analogous to the similar well-known ones 
for conformal transformations and conformal vector fields for $n=2$.

\subsection{Flat n-dimensional spacetime with parallel deformation direction}
Let us take flat $n$-dimensional Minkowski spacetime with Cartesian 
coordinates $\{x^{\mu}\}$, and let us pick up any covariantly constant 
null direction $\lh$. By adapting the coordinate system, we can 
always choose $\lh =du$ with $u\equiv (x^0 +x^1)/ \sqrt{2}$. Let us 
define  another null coordinate $v\equiv (x^1 -x^0)/ \sqrt{2}$ so that the 
line-element becomes
$$
ds^2=2dudv+\sum_{i}(dx^i)^2 , 
$$
where Latin small indices will take values $i,j,\dots =2,\dots ,n-1$.
In order to solve the Kerr-Schild equations 
(\ref{eq:KS1}-\ref{eq:KS2}) we can use a method similar to that of the 
previous two-dimensional case. Thus, (\ref{eq:KS2}) immediately leads 
to $\lie u=a(u)$ with $m=\dot a$. The contravariant form of (\ref{eq:KS2})
partly restricts further the form of $\xiv$ and can be used before 
attacking the first group of Kerr-Schild equations (\ref{eq:KS1}). 
Notice that the part of equations coming from the 2-planes $\{u,v\}$ 
are similar to the 2-dimensional case of the previous subsection 
with $f=0$, so that a part of the equations is already solved.
Then, the remaining part can be easily integrated
and we have the following result
\begin{prop} The KSVFs corresponding to a covariantly constant deformation
direction $\lh = du$ in flat spacetime are of the form
\begin{equation}
\xiv =   a(u) \frac{\partial}{\partial u}+ \left[b(u)-\dot{a}(u) v 
-\dot{c}_{i}(u) x^i\right] \frac{\partial}{\partial v}+
\left[c_{i}(u) +\epsilon_{ij}x^j\right]\frac{\partial}{\partial x^i},
\label{eq:KSv}
\end{equation}
where $a,b$ and $c_{i}$ are arbitrary functions of $u$, 
$\epsilon_{ij}=-\epsilon_{ji}$ are arbitrary constants, and the sum 
over repeated indices is to be understood.
Their associated deformation and metric gauge functions are given by
$$ m =\dot{a} , \hspace{1cm} h=\dot b -\ddot{a}\, v -\ddot{c}_{i} x^i .$$\N
\end{prop}

Thus, this Kerr-Schild algebra is uniquely characterized
by the {\em generating set} $\{a,b,c_{i},\epsilon_{ij}\}$ formed by 
$n$ arbitrary functions of $u$ and $(n-2)(n-3)/2$ arbitrary constants. 
Given that we are in the case of maximum degeneracy, in the sense that 
the metric has zero curvature and the deformation direction has 
vanishing covariant derivative, it seems plausible that the above is 
the maximum freedom one can have for a single Kerr-Schild algebra in 
general dimension $n$.

A direct evaluation leads to its {\em derived} algebra structure
\begin{prop} The Lie bracket $[\xiv , \vec \zeta ]$ of two KSVFs of 
type (\ref{eq:KSv}) characterized by the generating sets
$\{a,b,c_{i},\epsilon_{ij}\}$ and
$\{\tilde{a},\tilde{b},\tilde{c_{i}},\tilde{\epsilon}_{ij}\}$ 
respectively, is another KSVF with regard to $\lh =du$ whose
corresponding generating set reads
\begin{eqnarray}
\overline a & = & a \dot{\tilde{a}}- \dot{a}\tilde{a} \nonumber \\
\overline b & = & (a \tilde{b} - b \tilde{a})\dot{} +
\dot{c}_{i} \tilde{c}_{i}- c_{i} \dot{\tilde{c}}_{i} \nonumber \\
\overline{c}_{i} & = & a \dot{\tilde{c}}_{i} - \tilde{a}\dot{c}_{i}
+\tilde{\epsilon}_{ik} c_{k} -\epsilon_{ik} \tilde c_{k} 
\label{eq:gene} \\
\overline{\epsilon}_{ij} & = & \epsilon_{kj}\tilde{\epsilon}_{ik}-
\tilde{\epsilon}_{kj}\epsilon_{ik} .\nonumber
\end{eqnarray}\N
\end{prop}

Let us notice that the KSVFs (\ref{eq:KSv}) are Killing fields if
$h= 0$, which gives a Lie algebra of dimension
$$3+2(n-2)+\frac{(n-2)(n-3)}{2}=2+\frac{n(n-1)}{2} .
$$
A basis of this algebra is constituted by the $n$
translations together with the $(n-1)+(n-2)(n-3)/2$ rotations leaving
invariant the $(n-1)(n-2)/2$ two-planes containing $\lv $ and their orthogonal
vectors.

Expressions (\ref{eq:gene}) are useful to find special subalgebras.
For instance, one directly sees that the subalgebra of KSVFs defined by
$\epsilon_{ij}=0$ is an {\em ideal}, so that our Kerr-Schild algebra is not
{\em simple}.
This particular ideal is formed by Killing vectors, but 
this is not a general property. In the present case it is simply due to 
the particular form of flat spacetime. This will be clear from the
following results.
\begin{prop}
Any $g$ which is a Kerr-Schild transformed metric of flat spacetime 
with respect to a Minkowskian covariantly constant null deformation 
direction $\lh$ has the general solution (\ref{eq:KSv}) for the KSVFs with 
regard to $\lh$.
\end{prop}
\P This follows from Proposition \ref{same}, where the new metric gauge 
$\tilde{h}$ is given by $\tilde{h}= h + \lie H + 2mH$
(the other gauge function being invariant, $ \tilde{m} = m$.)\N

The spacetimes of the last Proposition have line-elements of type
$$
ds\sp{2} = 2 du dv + \sum_{i}(dx^i)^2 + 2 H(u,v,x^k)\, du\sp{2}
$$
and therefore they do {\it not} admit Killing vector fields in general. 
Thus, all the vector fields included in expression (\ref{eq:KSv}) are 
{\it proper} KSVFs for the generic metric above. In the particular 
case with $\partial H/\partial v =0$ we have the classical pp-waves 
metrics (in dimension $n$), see e.g. \cite{kramer,ek}.
These have a null Killing vector field along $\lv$.

\subsection{Spherically symmetric spacetime and deformation ($n=4$)}
Let us consider the most general spherically symmetric spacetime in 
the standard case of $n=4$. As is well known, there are only two 
independent {\it spherically symmetric} null directions, which are 
usually called the radial null directions. The congruences they define 
are always hypersurface-orthogonal, and thus we can select two null 
coordinates $u,v$ such that $du$ and $dv$ point along these two radial null 
directions. Completing the coordinate system with the angular 
variables $\theta,\varphi$, the most general line-element for such a 
spacetime can be written in the following simple form
\begin{equation}
ds^2=2e^fdudv+r^2d\Omega^2, \hspace{5mm} f(u,v), \hspace{5mm} r(u,v)
\label{ss}
\end{equation}
where the functions $f$ and $r$ are independent of the angular 
coordinates and $d\Omega^2$ is the line-element of the standard unit
2-sphere. The function $r$ has a clear geometrical meaning: $4\pi 
r^2$ is the area of the 2-dimensional spheres defined by constant 
values of $u$ and $v$, which are the orbits of the SO(3) group of motions.
Concerning the function $f$, it can be related to the invariantly 
defined {\it mass function} $M(u,v)$ by means of
$$
M(u,v)=\frac{r}{2}\left(1-2e^{-f}
\frac{\partial r}{\partial u}\frac{\partial r}{\partial v}\right).
$$
The selected coordinate system is preferable to the standard 
Schwarzschild coordinates for two reasons: it is clearly 
adapted to the null deformation directions we are going to study; and 
it allows to study spacetimes with event or Cauchy horizons. Thus, for 
instance, the maximal Kruskal extension of Schwarzschild spacetime can 
be described with a single coordinate system of the above type.

We have restricted ourselves to the case $n=4$ for the sake of 
simplicity and clarity, but it is evident that the analysis can be 
performed in general $n$ by simply substituting the metric of the
standard $(n-2)$-sphere for $d\Omega^2$ and SO($n-1$) for the isometry 
group. 

In order to get the general solution for KSVFs with respect to a 
radial null deformation direction let us take, without loss of 
generality, $\lh =du$. Then, as a first simple step we obtain the 
next result.
\begin{lem}
The Killing vector fields of the group SO(3) for
the general spherically symmetric spacetime (\ref{ss})
\begin{equation}
\vec J = c_1 \sin(\varphi + \varphi_{0})
\partial_{\theta} + \left( c_1 \cot \theta  \cos(\varphi + \varphi_{0})
+ c_2 \right) \pfi 
\label{kilss}
\end{equation}
are KSVFs with respect to $\lh =du$.
\end{lem}
\P It is enough to see that the vectors $\vec J$ leave $\Cl$ 
invariant, which is true because
$$
\pounds_{\vec J} \, \lh =\pounds_{\vec J} \, du =d(\pounds_{\vec J} u) =0,
$$
so that in fact we also have vanishing deformation gauge $m=0$ for 
the KSVFs (\ref{kilss}).\N

Therefore, the Lie algebra $\Kl$ has at least three free parameters 
$c_1,c_2,\varphi_{0}$. In order to complete the algebra $\Kl$ we can 
proceed as follows. We notice that the $\{u,v\}$-part of the 
line-element (\ref{ss}) is identical to the general 2-dimensional 
metric of subsection \ref{subsec:2dim}. Thus, our general solution 
will be of the form (\ref{2dim}) but restricted to satisfy the 
conditions coming from the angular part of the metric. This can be 
easily shown to lead to the intuitive condition\footnote{Of course, 
a similar reasoning can be applied to any metric in which there is a 
well-defined 2-dimensional subpart, such as the cases of decomposable 
spacetimes, warped products, or the general metric with an
isometry group acting on spacelike $(n-2)$-orbits.}
$$
\lie r =0
$$
which combined with (\ref{2dim}) gives the remaining solutions for $\Kl$. 
There appear several cases depending on the specific form of $r$. They 
are summarized as follows.
\begin{prop}
The KSVFs of the most general spherically symmetric metric with 
respect to a radial deformation direction $\lh =du$ are given by the 
Killing fields (\ref{kilss}) together with the following:
\begin{enumerate}
\item If $r=$const., the vector fields of the form (\ref{2dim}), with 
the deformation and metric gauges appearing in (\ref{2dimg}).
\item If $\partial r/\partial v =0$ but $r$ is not constant, the 
vector fields of the form
$$
\xiv =e^{-f}b(u)\partial_v
$$
where $b(u)$ is an arbitrary function and the gauges are
$$
m=0, \hspace{1cm} h= \dot b - b \frac{\partial f}{\partial u} .
$$
\item If $\partial r/\partial v \neq 0$, and {\em if}
$$
e^f = \frac{F(r)}{q(u)}\frac{\partial r}{\partial v}
$$
for some functions $F(r)$ and $q(u)$, the vector fields of type
$$
\xiv = q(u) \left(\partial_u -
\frac{\partial r/\partial u}{\partial r/\partial v}\partial_v\right)
$$
with the following gauges
\begin{equation}
m=\dot q , \hspace{1cm}
h=-\frac{F(r)}{q(u)}\frac{\partial r}{\partial v}\,\,
\frac{\partial}{\partial u}\!
\left(q(u)\frac{\partial r/\partial u}{\partial r/\partial v}\right).
\label{hss}
\end{equation}
\end{enumerate}
\N
\end{prop}
In case 1 the Lie algebra $\Kl$ is generated by two arbitrary 
functions and three constants, and in case 2 by one arbitrary function 
and the three constants. Notice that in case 3, and despite what it 
may seem, the solution depends on just four constants, as the 
function $q(u)$ appears explicitly in the metric (actually, this 
function could be set to a constant locally). In this last case, the 
fourth KSVF is proper in general, but there are some important cases 
in which it is in fact a Killing vector field. To find them,
from (\ref{hss}), it is necessary that
$$
q(u)\frac{\partial r/\partial u}{\partial r/\partial v}=p(v)
$$
where $p(v)$ is an arbitrary function of $v$. There are two cases 
now. If $p(v)=0$, then $r=r(v)$ and the line element reads simply
$$
ds^2 =2 \frac{\hat F (v)}{q(u)} dudv+r^2(v)d\Omega^2
$$
where $\hat F (v)\equiv F(r(v)) \dot r$. Notice that the 
$\{u,v\}$-part of the metric is flat, and the fourth KSVF is in fact
a {\it null} Killing vector field given by $\xiv = q(u) \partial_u$.
The second possibility is defined by $p(v)\neq 0$. In this case,
the function $r$ must have the form $r=r\left(P(v)+Q(u)\right)$ where
$\dot P =1/p$ and $\dot Q =1/q$ and the fourth KSVF which is a Killing reads
$$
\xiv = q(u)\partial_u - p(v)\partial_v .
$$
These spacetimes can be characterized by the property that the mass 
function depends only on $r$: $M=M(r)$. Then, it is easily seen that
the above KSVF is timelike or
spacelike depending on the sign of $1-2M/r$, as was to be expected.
This set of spacetimes includes {\it all} spherically symmetric metrics 
with a {\it static} region, such as Minkowski spacetime, 
Schwarzschild vacuum solution and its maximal Kruskal extension, 
Schwarzschild interior solution, all static spherically symmetric 
perfect fluids, Reissner-Nordstr\"om charged solution and its maximal 
extension, Einstein static universe, de Sitter spacetime, and many 
others.

As a simple but illustrative example of a physical case in which the 
fourth KSVF of case 3 is proper, we can take the Vaidya radiative 
solution \cite{Vaidya}, which is a Kerr-Schild transformed metric of 
flat spacetime, as is known \cite{null,taub}. In our coordinates, the 
Vaidya solution is given by
$$
F(r)=C=\mbox{const.}, \hspace{1cm} M=M(u), \hspace{1cm}
\frac{\partial r}{\partial u} 
=\frac{1}{2q(u)}\left(1-\frac{2M(u)}{r}\right)
$$
where the mass $M(u)$ is an arbitrary funtion of $u$ (the Schwarzschild 
metric is contained as the case $\dot M =0$.) The proper KSVF reads
$$
\xiv = q(u)\partial_u +
\frac{e^{-f}}{2}\left(1-\frac{2M(u)}{r}\right)\partial_v
$$
and its metric gauge is
$$
h=\frac{C}{q(u)}\,\frac{\dot M}{r} .
$$

\subsection{Flat spacetime with cylindrical deformation direction 
($n=4$)}
In all previous cases, the congruence $\Cl$ defined by the deformation 
direction $\lh$ was irrotational (and therefore geodesic) and 
shear-free. Now, we are going to present a simple case of a {\it 
shearing} deformation direction, given by a cylindrical null 
direction in Minkowski. Again, for the sake of simplicity we assume 
$n=4$, but the results can be straightforwardly generalized to any $n$.

By using a classical cylindrical coordinate system
$\{t, \rho, \varphi, z\}$, a cylindrical null
direction in Minkowski spacetime is given by
$ \lh = d(t+\rho)$. We can select advanced and retarded null coordinates,
$u ={(t+\rho) / {\sqrt 2}}$ and $ v = {(t - \rho) / {\sqrt 2}} $, so that
$ \lh $ writes now as $\lh = du $. The Kerr-Schild equations for this null
direction can also be explicitly integrated, their general solution
being
$$ \xiv = (c_0\,u + c_1)\partial_{u} + (2c_0\,u - c_0\,v + c_1)\partial_{v} +
(c_0\,\varphi + c_2)\pfi + c_3\partial_{z}, $$
where $c_{\alpha}$ are four arbitrary
constants. For any value of them one has $h = - 2c_0$ and $m =c_0$
for the gauges.

This case presents an interesting feature, unusual in General
Relativity: the above general solution corresponds to {\em proper}
KSVFs but, due to the presence of the term $\varphi \pfi$,
they are {\em local} vector fields (they would be
bivalued after a complete revolution). Only when $ c_0 = 0$ they
become {\em global}, but then they reduce to Killing vector fields with
$m=0$.
Denoting by $\xiv_{\alpha}$ the KSVF obtained from the above general
solution by setting the constants equal to zero except for the 
$\alpha$-th, we have
\begin{prop}
For a cylindrical Kerr-Schild deformation in flat spacetime,
proper KSVFs are necessarily {\em local},
and form a four-dimensional Lie algebra, their non vanishing
brakets being
$$[\xiv _0,\xiv _1] = - c_0 \, \xiv _1 \ \ , \ \  [\xiv _0,\xiv _2] = -
c_0 \, \xiv _2 .$$
Global KSVFs are necessarily Killing vector fileds, and reduce
to the static cylindrical symmetry which is Abelian (three-dimensional
translation abstract group).\N
\end{prop}

\section{Conclusion}

As we have seen, the notion of Kerr-Schild vector fields seems to be 
meaningful and, in fact, it leads to a structure richer than that of
the classical Killing or conformal fields. As we have seen, one can 
define the set of all KSVFs in the spacetime and give the general 
equations for them, independently of the deformation direction $\lh$. 
However, this set has not even the structure of a vector space in 
general. Nevertheless, the KSVFs with respect to $\lh$ constitute a 
Lie algebra. These are generically 
finite-dimensional, even though they can be of infinite 
dimension in some particular cases which are of relevance. So far,
one knows very little about the structure of these Kerr-Schild infinite
algebras. We have shown that they are not simple, but a
formal proof that they do not admit Abelian ideals is lacking, as 
well as the characterization of the possible Abelian subalgebras.

Many questions remain open. For instance, the necessary and sufficient
integrability conditions of the Kerr-Schild equations, or the 
construction of a complete set of geometrical objects which are 
invariant under Kerr-Schild transformations, and under KSVFs. In this 
sense, we have proved the result that any two 2-dimensional metrics are
Kerr-Schild transformed of each other, and of the flat metric, with respect
to any of the two possible null deformation directions. This is 
analogous to the result that establishes the conformally flat character 
of any 2-dimensional metric. But the corresponding result for general 
dimension $n$ is still open. On the other hand, and analogously to the case
of Killing fields, which become conformal fields by a conformal transformation,
we have seen that Killing fields leaving invariant a null deformation direction
$\lh$ become KSVFs by a Kerr-Schild transformation. However, we do not
know the analogue for KSVFs of the Defrise-Carter theorem \cite{nota6}
for conformal transformations, namely, how to control the number of
KSVFs that may become Killing fields by a suitable Kerr-Schild
transformation. Some of these open problems will be
considered elsewhere \cite{nota5}.

\section*{Acknowledgements}
JMMS is grateful to Marc Mars and Ra\"ul Vera for reading the manuscript and
for some valuable comments. SRH wishes to thank the Comissionat de
Recerca i Universitats for finantial support.
JMMS wishes to thank financial support from the Basque Country 
University under project number UPV 172.310-G02/99.


\begin{thebibliography}{99}
\bibitem{nota4}{Riemannian and Symplectic manifolds show that strict
or conformal invariance of symmetric and antisymmetric regular
tensors is respectively finite and infinite dimensional. It is
interesting to observe that the non-homogeneous character of the equations
in the Kerr-Schild case is responsible for the possibility of having infinite
dimensions, despite the symmetry and the regular character of
the tensor $g$ involved.}
\end{thebibliography}
\end{document}